\documentclass[runningheads]{llncs}
\usepackage{graphicx}
\usepackage[nice]{nicefrac}
\usepackage{amsmath}
\usepackage{amsfonts}
\usepackage{mathtools}

\begin{document}

\title{Stacked Hourglass Network with a Multi-level Attention Mechanism: Where to Look for Intervertebral Disc Labeling}

\author{Reza Azad\inst{1} \and
Lucas Rouhier\inst{2} \and
Julien Cohen-Adad\inst{2, 3, 4}}

\authorrunning{R. Azad et al.}
\titlerunning{Stacked Hourglass Network with a Multi-level Attention Mechanism}
%
\institute{Sharif University of Technology, Tehran, Iran, rezazad68@gmail.com \and
NeuroPoly Lab, Institute of Biomedical Engineering, Polytechnique Montreal, Canada, lucasrouhier@gmail.com, jcohen@polymtl.ca \and
Mila, Quebec AI Institute, Canada \and
Functional Neuroimaging Unit, CRIUGM, University of Montreal, Montreal, Canada}

\maketitle              
\begin{abstract}
Labeling vertebral discs from MRI scans is important for the proper diagnosis of spinal related diseases, including multiple sclerosis, amyotrophic lateral sclerosis, degenerative cervical myelopathy and cancer. Automatic labeling of the vertebral discs in MRI data is a difficult task because of the similarity between discs and bone area, the variability in the geometry of the spine and surrounding tissues across individuals, and the variability across scans (manufacturers, pulse sequence, image contrast, resolution and artefacts). In previous studies, vertebral disc labeling is often done after a disc detection step and mostly fails when the localization algorithm misses discs or has false positive detection. In this work, we aim to mitigate this problem by reformulating the semantic vertebral disc labeling using the pose estimation technique. To do so, we propose a stacked hourglass network with multi-level attention mechanism to jointly learn intervertebral disc position and their skeleton structure. The proposed deep learning model takes into account the strength of semantic segmentation and pose estimation technique to handle the missing area and false positive detection. To further improve the performance of the proposed method, we propose a skeleton-based search space to reduce false positive detection. The proposed method evaluated on spine generic public multi-center dataset and demonstrated better performance comparing to previous work, on both T1w and T2w contrasts. The method is implemented in ivadomed (https://ivadomed.org).

\keywords{intervertebral disc labeling \and
spine generic database \and pose estimation \and deep learning.}
\end{abstract}
\section{Introduction}
\begin{figure}[ht]
\label{fig:spine_cloumns}
	\centering
	\begin{tabular}{cc}
		\includegraphics[width=0.60\textwidth]{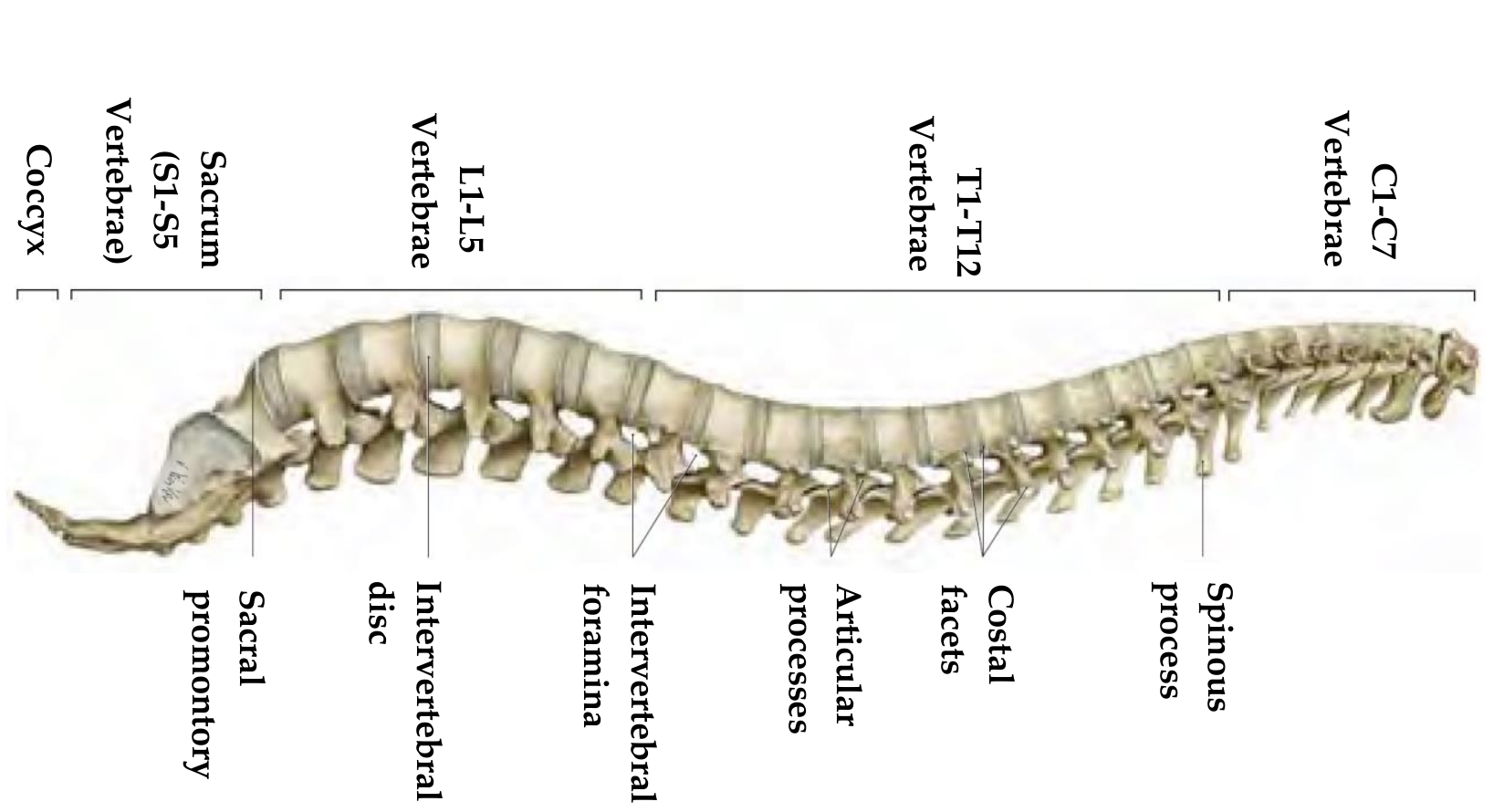}&
	\end{tabular}
	\caption{Human vertebral column \cite{vertebralcolumn}}.
\end{figure}

The human spine consists of four connected regions namely: cervical, thoracic, lumbar, and sacral vertebrae (Figure \ref{fig:spine_cloumns}).  The vertebrae in each region perform unique functionality, including protection of the spinal cord, load breathing, and etc. As shown in Figure \ref{fig:spine_cloumns}, intervertebral discs lie between adjacent vertebrae and connect the vertebrae column. Any injury in the vertebral disc may result in back pain or sensation in different parts of the human body. Vertebral injury usually comes from excessive strain or trauma to the spine. Thus, analyzing the intervertebral disc location and its shape is an important part of diagnosis. 
The analysis starts with the detection of intervertebal discs, which is a tedious job. To automate the intervertebal disc detection process, several approaches have been proposed in previous studies. A local descriptor based approach has been utilized by \cite{gros2018automatic} to detect the intervertebral disc C2/C3. This approach uses the local descriptor to find the mutual information between the image from the patient and the template by looking at region that has the minimum distance to the spine template. Even though this handcrafted approach produces good results in general, its performances are greatly reduced when the images differs too much from the template. To overcome the limitation of handcrafted approaches, deep learning based models are used to perform robust intervertebral disc labeling. In \cite{chen2019vertebrae} the author proposed a 3D CNN model to perform the 3D segmentation on MRI data and retrieve the vertebral disc location. Cai et al. \cite{cai2015multi} also utilized a 3D Deformable Hierarchical Model to extract the vertebral disc location on 3D space. In an another example, the Count-ception model is trained on 2D MRI sagittal slices to perform vertebral discs detection \cite{rouhier2020spine}. The exact location of the vertebral discs in 2D space are extracted using local maximum technique on top of the prediction masks. The mains drawback of the deep segmentation approaches is the false positive rate. Therefore, it requires extensive post-processing and it often fails to retrieve the exact location of the disc. In this work, we aim to overcome the limitations of literature work by taking into account the importance of skeleton structure in intervertebral disc labeling. The main idea is to include the structural information of the intervertebral discs to increase True Positive (TP) rate while reducing the False Negative (FN) detection. We re-formulate the problem using pose estimation technique, to take into account the skeleton structure between intervertebral discs in loss function. To the best of our knowledge, this is the first attempt to label intervertebral discs using pose estimation technique. Our contribution can be summarized as follow:

\begin{itemize}
    \item Adapting the pose estimation approach for intervertebral disc labeling.
    \item Scaling the representation space by multi-level attention mechanism.
    \item Skeleton based post-processing approach to reduce the false-positive rate.
    \item State-of-the-art results on the public data set.
    \item Publicly-available implementation source code \cite{SHNIDLcode} 
    
\end{itemize}

\section{Literature review}
Automatic intervertebral disc labeling is a crucial task in medical image analysis.  The spine can be considered as an anatomical landmark that runs along the length of the neck and trunk of the human upper body. In recent decades, many approaches have been proposed in this field of research. 
Like other tasks in computer vision, we can divide these approaches in two groups: handcrafted-based methods and deep learning-based approaches. Early studies in intervertebral discs labeling were based on handcrafted-based features. Glocker et al. \cite{glocker2012automatic} employed regression forests and hidden Markov model (HMM) to detect and localize intervertebral discs from CT scan images. Kim et al. \cite{kim2017vertebrae} proposed a Sphere Surface Expansion method and iterative optimization framework by incorporating local appearance features with global translational symmetry and local reflection symmetry features.  
Zhang et al. \cite{zhan2012robust} utilized a local articulated model to effectively model the spatial relations across vertebrae and discs. The authors derived disk locations from a cloud of responses from disc detectors which is robust to sporadic voxel-level errors. 

Ullmann et al. \cite{ullmann2014automatic} proposed a 3D-based analysis method, which supports both T1-weighted and T2-weighted contrasts. A template is utilized for detecting the intervertebral disc location. This template was computed from a collection of vertebral distances along the cervical, thoracic, and lumbar spine in adult humans. 
A method for detecting the spinal cord centerline on MRI volumes was proposed by Gros et al. \cite{gros2018automatic}. The optimization procedure of that method aimed at striking a balance between a probabilistic localization map of the spinal cord center point and the overall spatial consistency of the spinal cord centerline. The authors also employed a post-processing step to split brain and spine region.

The handcrafted-based features suffers from overfitting problem due to the over-designed framework. Like other field of research in computer vision, in recent years, deep learning-based approaches have been proposed for detection of spine vertebrae. Rouhier et al. \cite{rouhier2020spine} combine a Fully Convolutional Network (FCN) with inception modules to localize intervertebral discs from MRI data. An FCN was also utilized by Suzani et al. \cite{suzani2015fast} to predict the relative distance from the voxel to each vertebral centroid. 
Forsberg et al. \cite{forsberg2017detection} employed two separate detection and labeling pipelines, for the lumbar and the cervical cases. Each pipeline consists of two convolutional neural networks (CNNs) for detection of potential vertebrae (one general lumbar/cervical vertebra detector and one specific S1/C2 vertebra detector, respectively), followed by a parts-based graphical model for false positive removal and subsequent labeling.

A deep convolutional neural network was exploited by Chen et al. \cite{chen2015automatic} to identify the vertebrae type based on the centroid proposals generated from the random forest classifier. 
Transformed Deep Convolution Network (TDCN) was introduced by Cai et al. \cite{cai2016multi} for multi-modal intervertebral disc recognition. The image features from different modalities are fused unsupervisely, and the pose of vertebra is then automatically rectified.  Windsor et al. \cite{Windsor2020ACA} proposed a CNN model for detection and localization of vertebrae from MRI images. The model is uses the image-to-image translation technique to label the intervertebral disc. 

Lu et al. \cite{lu2018deepspine}, utilized a  natural-language-processing approach to extract level-by-level ground-truth labels from free-text radiology reports, and then employed a U-Net architecture combined with a spine-curve fitting method for intervertebral segmentation and disc-level localization. A multi-input, multi-task, and multi-class CNN is then exploited for central canal and foraminal stenosis grading. A U-Net like architecture is proposed by Yang et al. \cite{yang2017automatic} to directly model the vertebrae centroids as a 27-class segmentation-like problem (26 vertebrae types and background) 
An FCN-based approach was employed by Chen et al. \cite{chen2019vertebrae} for vertebrae localization. The authors then propose a post-processing method to reduce the dimension of the localization, and constrain the search space for the vertebrae centroids with a hidden Markov model. The reviewed deep learning-based methods does not consider the geometrical information in the learning process, hence, suffers from FP and FN detection. We aim to overcome this limitation using the pose estimation technique. 

\section{Proposed Method}
\begin{figure}
\includegraphics[width=\textwidth]{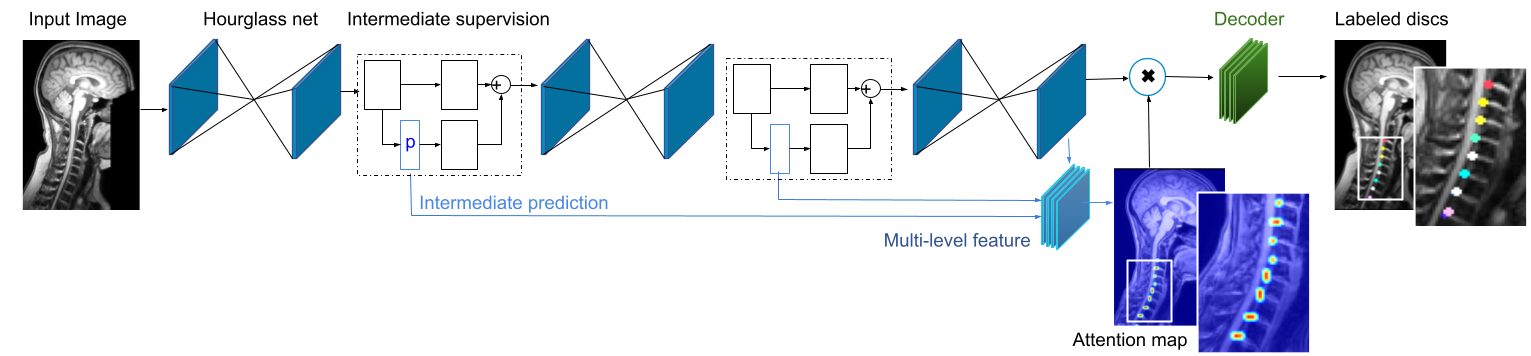}
\caption{Stacked hourglass network with an attention mechanism. The model considers the loss function between each hourglass prediction and ground truth mask (intermediate supervision). Further it feds the intermediate representation into an attention layer to produce the attention map. The attention map guides the decoder layer to focus on the intervertebral disc. } \label{fig:proposed_method}
\end{figure}

General diagram of the proposed method is shown in Figure \ref{fig:proposed_method}. The first step in the proposed method is to pre-process the input data for the model. The model utilizes the pose estimation method with attention mechanism to learn intervertebral disc position. In the next subsections we will discuss these steps.  

\subsection{Pre-processing}
As done in previous studies \cite{rouhier2020spine}, we extract the average of $6$ sagittal slices (centered in the middle slice) as a data sample for each subject. We normalize each image to be in range $[0, 1]$ to reduce the effect of data variation.
In order to prepare the ground truth data for the training process, first, we extract the intervertebral disc position (single pixel) from the ground truth data then we convolve the image with a Gaussian kernel to generate a smooth ground truth with increased target size (radius 10). We repeat this process for each intervertebral disc separately to produce $V$ channel ground truth, where $V$ is the number of intervertebral discs. Since the Spine Generic dataset consists of samples with variable number of intervertebral discs (between 7-11), we extract 11 intervertebral discs for each subject. For any missing intervertebal disc we consider unknown position and eliminate its effect on the training process by simply filtering out with the visibility flag on the loss function.

\subsection{Proposed model}
As shown in Figure \ref{fig:proposed_method}, the stacked hourglass network \cite{newell2016stacked} learns the object pose using (N-1) intermediate (shown in \ref{fig:proposed_method} as intermediate prediction) prediction and one final prediction. Thus, it takes into account the multi-level representation in terms of the N stacked hourglass network. To further improve the power of representation space, we propose to use a multi-level attention mechanism. To this end, an intermediate representation generated by each hourglass network (shown in \ref{fig:proposed_method} with $p$) is concatenated to form a multi-level representation. This representation can be seen as a collective knowledge that is extracted from a different level of the network with various scales, thus, using this collective knowledge as a supervisory signal to calibrate the final representation can result in better representation. To include this supervisor signal, we stack all the intermediate representations. This stacked representation is fed to the attention block (series of point-wise convolution with sigmoid activation) to generate a single channel attention mechanism. We multiply this attention channel with the final representation to re-calibrate the representation space and teach the model to pay more attention to the disc location. This attention mechanism reduces the FP rates. We train the model using the sum of MSE loss (equation \ref{eq:trainingloss}) between the predicted mask $y'$ and the ground truth mask $y$. In equation \ref{eq:trainingloss}, N shows the number of pixels in the ground truth mask. 

\begin{equation}
\mathrm{loss}=\frac{1}{V \times N} \sum_{i=1}^{V}\sum_{j=1}^{N}\left(y_{j}-\hat{y}_{j}\right)^{2}
\label{eq:trainingloss}
\end{equation}

\subsection{Post-processing}
\begin{figure}
\centering
\includegraphics[width=0.7\textwidth]{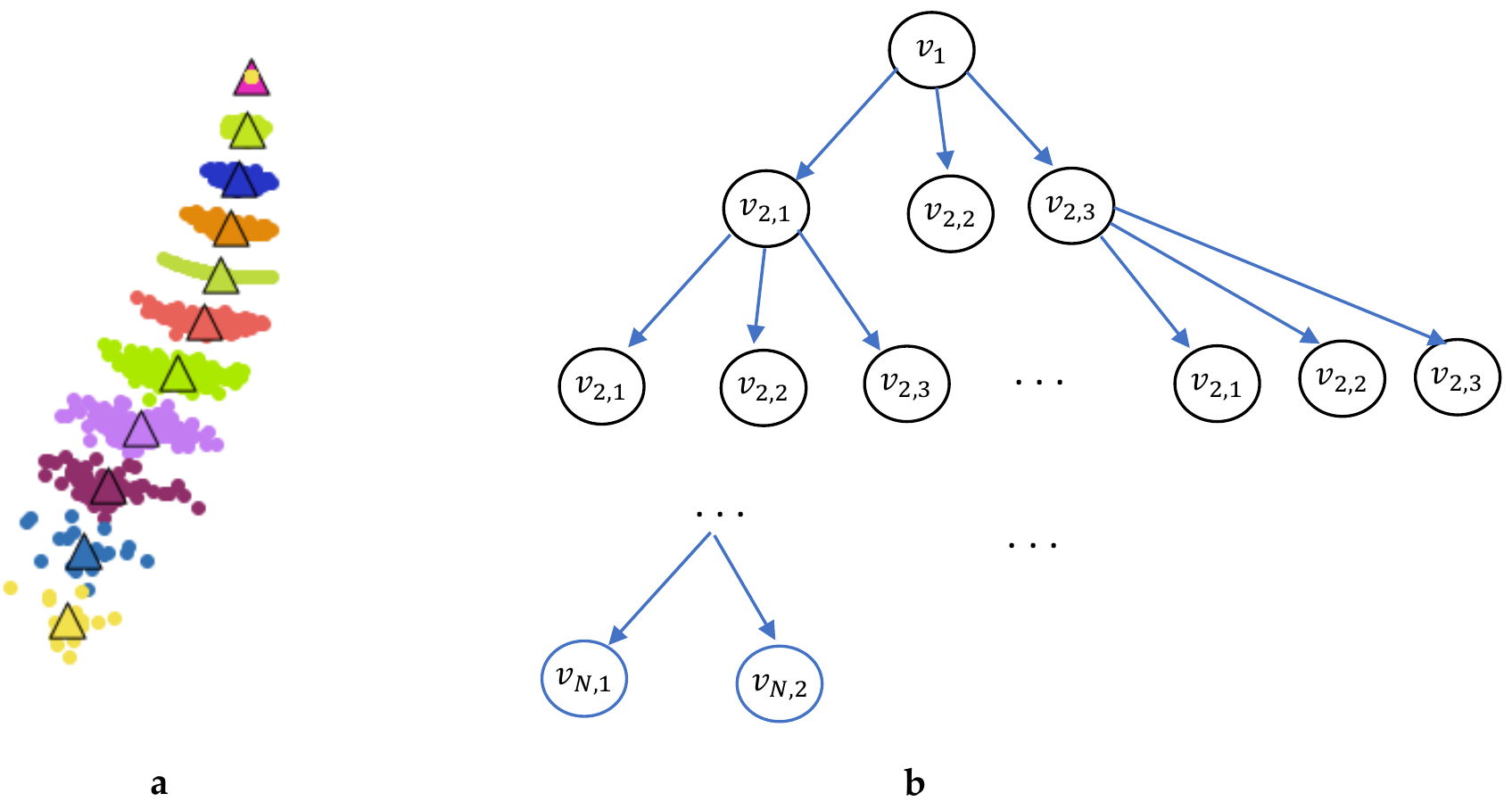}
\caption{(a): Vertebral disc skeleton structure (b): Skeleton search tree, where each level of the tree contains candidates for invertebrate $v_i$. Path from root to leaf shows one possible combination of detected candidates to form a $S'$.} \label{fig:3}
\end{figure}

Even though the proposed network learns the intervertebral discs with high accuracy, the predicted mask needs to be post-processed to further reduce the FP rate. In order to fine-tune the predicted result we propose a skeleton-based approach. In this approach, we create a general skeleton model based on the training set. 
To this end, for all subjects in the training set, we extract the intervertebral disc location, then we calculate the distance from each intervertebral disc ($v^i$) to the first intervertebral ($v^1$) disc. Which shows the relational structure between intervertebral discs. To normalize this representation, we shift the $v^1$ to the world coordinate $(0,0)$ then we normalize all the relational distances by dividing them to the distance from $v^1$ to $v^5$. Figure \ref{fig:3} (a) shows the structure of the extracted skeleton from the training set. Each intervertebral disc $v^i$ (shown by triangle) is calculated based on the average of all subject's intervertebral disc location $v^i$. Using the generated skeleton, we define the skeleton model S as \ref{eq:skeleton}:

\begin{equation}
S=\operatorname{set}\left\{v^{i}\right\}, i=1,2, \ldots V, v^{i}=(x, y)    
\label{eq:skeleton}
\end{equation}

Skeleton S consists of V intervertebral discs with relational 2D position (x, y). On the test time, for each predicted mask, we create the search tree, where each node in this tree shows one possible combination of ordered intervertebral disc location, as illustrated in Figure \ref{fig:3} (b). 
For each path of tree (from leaf to node which forms a S’) we calculate the error function between the general skeleton S and the predicted skeleton S’ using equation \ref{eq:errorfunction1}. Please note that for each intervertebral disc ($v_i$) we have several candidates, thus, ($v_i,c$) shows the $c_{th}$ candidate of $v_i$. In our equation S’ with minimum error is the solution. We use flag $\delta$ in equation \ref{eq:errorfunction1} to represent that candidate availability. Hence, in case the algorithm misses any intervertebral disc, it will not affect the error function.

\begin{equation}
\operatorname{error}\left(S, S^{\prime}\right)=\sum_{j=1}^{N} d\left(v^{j}, v^{j^{\prime}}\right), \quad d\left(v^{j}, v^{j^{\prime}}\right)=\sqrt{\sum_{i=1}^{n}\delta\left(v_{i}^{j}-v_{i}^{\prime j}\right)^{2}}
\label{eq:errorfunction1}
\end{equation}

\section{Experimental Results}
To evaluate the performance of the proposed method we consider the Spine Generic Dataset \cite{spinedataset}. This dataset contains both T1w and T2w contrasts for each subject. Images were acquired in 42 different centers world wide. The dataset contains large sample variation in term of image quality, scale and imaging devices, hence, exhibits a challenging benchmark for intervertebral disc labeling.

\subsection{Metrics}
To demonstrate the performance of the proposed method and compare with the literature work, we consider different comparison metrics.
We include the L2 distance between each predicted intervertebral disc position and the ground truth along the superior-inferior axis to calculate the preciseness of the prediction. We further use the False Positive Rate (FPR) and False Negative Rate (FNR) metrics. The FPR counts the number of prediction which are at least 5 mm away from the ground truth positions. Similarly the FNR counts the number of prediction where the ground truth has at least 5mm distance from the predicted intervertebral disc's real position.

\subsection{Comparison Results}
We train the proposed model for 150 epochs using Adam optimization with learning rate $0.00025$ and batch size $4$. In our experimental results we achieved best results on the validation set using 2 stacks.  The  implementation and model training was done in ivadomed \cite{Gros2021} and the method can readily be used via the Spinal Cord Toolbox \cite{de2017sct}. Comparison results of the proposed method with the literature work on the test set is provided in table 1. We use the same setting as explained in \cite{rouhier2020spine} to compare our method with the literature work. As shown in table 1, the proposed method outperformed literature work in both T1w and T2w modalities. On T1w modality, the proposed method has better results than the template matching technique on all metrics. Although the counting method on T1w modality has a slightly lower average distance to the target, our proposed method has a lower standard deviation value and statistically it is more reliable for retrieving intervertebral disc location. The efficiency of the proposed method is clearer on the T2 contrast, where the proposed method outperforms the Count-ception and template matching methods.

\begin{table}

\centering
	\caption{Intervertebral disc labeling results on the spine generic public dataset.}\label{tab1}
	\resizebox{\textwidth}{!}{%
	\begin{tabular}{|p{0.3\textwidth}|*{2}{p{0.3\textwidth}|p{0.15\textwidth}|p{0.15\textwidth}|}}
		\hline
		\multicolumn{1}{|c|}{\textbf{Method}} &
		\multicolumn{3}{|c|}{\textbf{T1}} &
		\multicolumn{3}{|c|}{\textbf{T2}} \\
		\hline
		  & Distance to target (mm)&	FNR (\%)&	FPR (\%)&	Distance to target (mm)&	FNR (\%)&	FPR (\%)\\
		\hline
	    Template Matching  & 1.97($\pm$ 4.08)&	8.1& {2.53}&	2.05($\pm$ 3.21)&	11.1 & 2.11\\
		\hline
	    Counting model \cite{rouhier2020spine}  & 1.03($\pm$ 2.81)&	4.24& {0.9}&	1.78($\pm$ 2.64)&	3.88 & 1.5\\
		\hline	
	    Proposed Method (without attention) & 1.39($\pm$ 2.38)&	1.3& {0.0}&	1.31($\pm$ 3.34)&	1.5 & 0.0\\
		\hline			
		Proposed Method& \textbf{1.32($\pm $ 1.33)}&\textbf{0.32}&0.0&	\textbf{1.31($\pm $2.79}&\textbf{	1.2}& 0.6\\
		\hline
	\end{tabular}
	}
\end{table}

It is worth mentioning that the counting method \cite{rouhier2020spine} applies image straightening technique on the pre-processing step to extract the spinal cord area. This technique crops the spinal region from the input MRI image ad simplifies the data sample. However, in our proposed method we eliminated this step to learn the intervertebral disc labeling without such pre-processing requirement. In addition, the counting method doesn't consider the relation between intervertebral discs and simply applies the region based segmentation technique to extract the intervertebral disc location. Thus, it has high false negative rate (FNR). On the other hand, the proposed method takes into account the skeleton information to eliminate the FNR samples. In our post-processing method, to eliminate the FN samples and reduce the search burden we provide the number of desired intervertebral disc as a meta data. Thus, this metadata helps the search tree to create a less depth and retrieve the intervertebral disc location fast.

\begin{figure}
\centering
\includegraphics[width=0.7\textwidth]{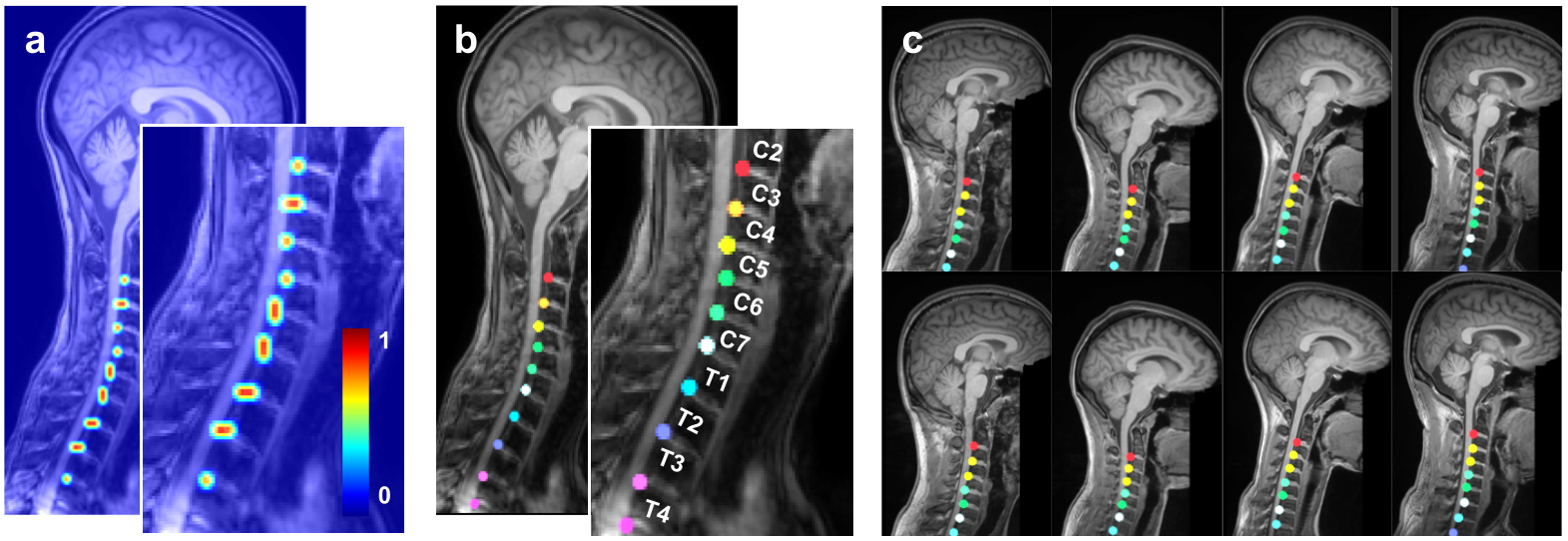}
\caption{(a): Attention map visualization. (b): Corresponding ground truth mask. (c): Prediction results in four representative T1w images. The top row shows the ground truths, the bottom row shows the predictions. } \label{fig:4}
\end{figure}

\subsection{Effect of attention mechanism}
The proposed method utilizes the attention mechanism to re-calibrate the representation space and guide the model to focus on target location. To analyse the effect of the attention mechanism we trained the model with and without attention mechanism. The comparison results demonstrated in table 1, in which the attention mechanism in both T1w and T2w modalities increases the model performance. To visualize the effect of attention mechanism inside the network, we depicted a sample attention maps on the input images, Figure \ref{fig:4} (b).

\section{Conclusion}
In this work we formulated the intervertebral disc labeling using pose estimation technique. The proposed method learns the structural information between intervertebal discs to recover the TP locations. The proposed method uses the strength of attention mechanism to re-calibrate the representation space in way to focus more on interverterbral disc area. We further proposed a skeleton-based post processing approach to eliminate the FP and FN detection.

%
%
%

 \bibliographystyle{splncs04}
 \bibliography{Ref}





\end{document}